\documentclass[seceq,epsf]{ptptex}

\usepackage{graphicx}

\newcommand{\gsim}
{\mathrel{\mbox{\raisebox{-1.0ex}
    {$\stackrel{\displaystyle >}{\displaystyle \sim}$}}}}
\newcommand{\lsim}
{\mathrel{\mbox{\raisebox{-1.0ex}
    {$\stackrel{\displaystyle <}{\displaystyle \sim}$}}}}

\notypesetlogo                       

\title{
Phase Structure of Hot and/or Dense QCD \\ in the Schwinger-Dyson Approach%
}

\author{
Satoshi \textsc{Takagi}%
}

\inst{%
Department of Physics, Nagoya University, Nagoya 464-8602, Japan

}

\abst{
We investigate the phase structure of hot
and/or dense QCD with massless 2-flavors
using the Schwinger-Dyson equation (SDE) with the improved ladder
approximation in the Landau gauge.
We examine the effect of the antiquark contribution 
and find that setting the antiquark Majorana mass 
equal to the quark one is a good approximation in the medium density region.
The imaginary part of the Dirac mass has an important influence on 
the chiral phase transition, in particular on the tricritical 
point.
}

\begin{document}

\maketitle

\section{Introduction}
\label{Introduction}

The exploration of the QCD phase diagram, 
including the color superconducting phase, 
is an important subject 
for the purpose of studying 
not only the dynamics of QCD
but also its phenomenological applications in cosmology, the
astrophysics of neutron stars and the physics of heavy ion 
collisions.~\cite{RK}
The phase structure of QCD at non-zero temperature ($T$) with
zero chemical potential has been extensively studied with lattice simulations,
but the simulations at finite chemical potential ($\mu$) has begun only
recently and these simulations still involve large errors
[see, e.g., Refs.~\citen{Karsch:2003jg} 
and references cited therein].
Thus, it is important to investigate 
the phase structure of QCD in the finite $T$
and/or $\mu$ region by many other approaches. 

In various non-perturbative approaches, the approach based on  
the Schwinger-Dyson Equation (SDE) is one of the most powerful tools
[see references cited in Ref.~\citen{takagi}]. 
{}From the SDE with a suitable running coupling, 
the high energy behavior of the mass function 
is shown to be consistent with the result derived from QCD 
with the operator product expansion and the renormalization group equation. 
Furthermore, the SDE  
include the effect of the long range force mediated by
the magnetic mode of the gluon, 
which may have a substantial effect, 
even in the intermediate chemical potential region, 
as in the high density region.~\cite{Son,Hong,tsch,pis}
In the SDE analysis,
the phase structure of QCD in the finite $T$ and/or $\mu$ region 
have been investigated, concentrating on 
the chiral symmetry restoration.~\cite{Tani,Ha,Bar,Kir1}

In Ref.~\citen{takagi} 
we solved the coupled SDE for the Majorana masses of the quark
($\Delta^-$)
and antiquark ($\Delta^+$) separately from the SDE for the Dirac mass
($B$) 
in the low and intermediate temperature and chemical potential region. 
The true vacuum is determined by comparing the values 
of the effective potential at these solutions.
Here we report the results in Ref.~\citen{takagi} and add a figure 
in order to emphasize the effect of the imaginary part of the Dirac mass.

\section{Phase structure}
\label{Phase Structure}

\begin{figure}[hb]
 \centerline{\includegraphics[width=12cm]{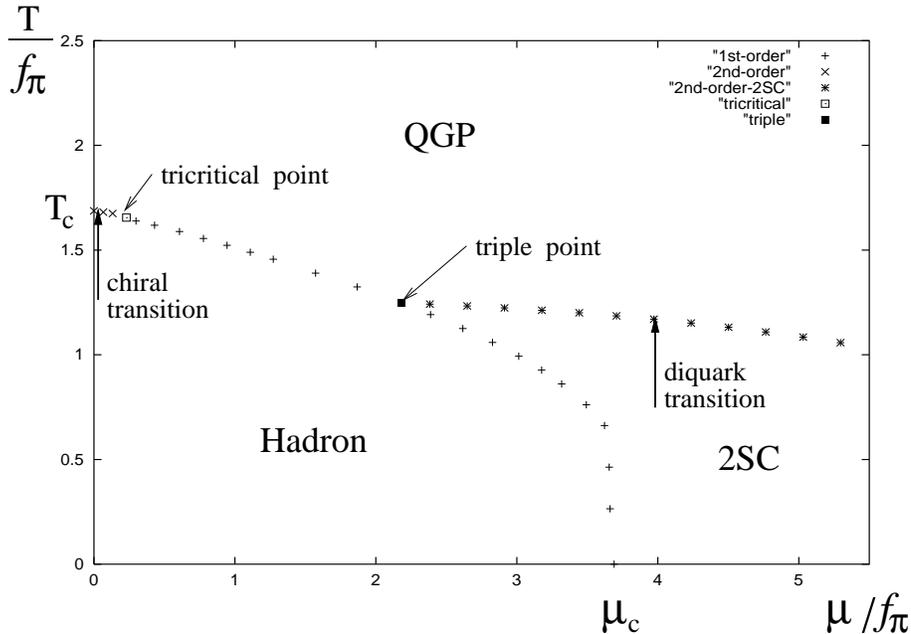}}
 \caption[]{Phase diagram
 for $0\leq{T}/f_\pi\leq{2.5}$ and $0\leq\mu/f_\pi\leq{5.5}$. 
The symbol $\times$ denotes the second order phase transtion
 between the hadron phase and the QGP phase,
$+$ the first order phase transition between the hadron phase
 and the QGP phase or the 2SC phase,
 and $\times\hspace{-0.75em}+$ the second order phase
 transition between the 2SC phase and the QGP phase.
 $T_c$=147 \mbox{MeV}, $\mu_c$=325 \mbox{MeV},
 and ($T$, $\mu$)=(146, 20) \mbox{MeV} at the tricritical point,
 where the second order phase transition changes to 
 first order. The value of the pion decay constant $f_\pi$ is 88 \mbox{MeV}. 
 }
 \label{phase-structure}
\end{figure}

The phase diagram obtained in Ref.~\citen{takagi} 
is shown in Fig.~\ref{phase-structure}.
We consider the possibility of three phases:
the hadron phase, the two-flavor color superconducting (2SC) 
phase~\cite{Al,Rapp:1997zu} and the quark-gluon plasma (QGP) phase.
The value of the pion decay constant $f_\pi$ is 88 MeV.
There are two notable features in Fig.~\ref{phase-structure}.
One is the position of the tricritical point:
$(T, \mu)=(146, 20)\,\mbox{MeV}$
(indicated by {\small $\square$} in Fig.~\ref{phase-structure}).
The origin of this feature in the SDE is discussed in the next section.
The other is the high critical temperature of the color phase transtion
(2SC$\leftrightarrow$QGP). It is around $100\,\mbox{MeV}$
for any $\mu$ satisfying $\mu\lsim500\,\mbox{MeV}$,
and this value is larger than those obtained in 
models based on the contact 4-Fermi 
interaction[see, e.g., Ref.~\citen{Be}].
This increase of the critical temperature may be 
caused by the long range force.

We examined the antiquark contribution of the 2SC phase in Ref.~\citen{takagi}.
Our results show that the antiquark Majorana mass gap, $\Delta^+$,
is comparable to the quark one, $\Delta^-$,
in all the chemical potential regions that we studied:
$1 > \Delta^+/\Delta^- \gsim 0.85$ for 
$0 < \mu \lsim 500\,\mbox{MeV}$~\cite{takagi}.
However, we find that $\Delta^+$ can not be generated 
in the SDE without $\Delta^-$~\cite{takagi}.
Therefore the anti-quark gap $\Delta^+$ is generated radiatively 
by the quark gap $\Delta^-$ as shown 
in the weak coupling limit~\cite{Hong:2000ng}.

\section{Effect of the imaginary part of the Dirac mass}
\label{imaginary part}

\begin{figure}[htb]
 \centerline{\includegraphics[width=12cm]{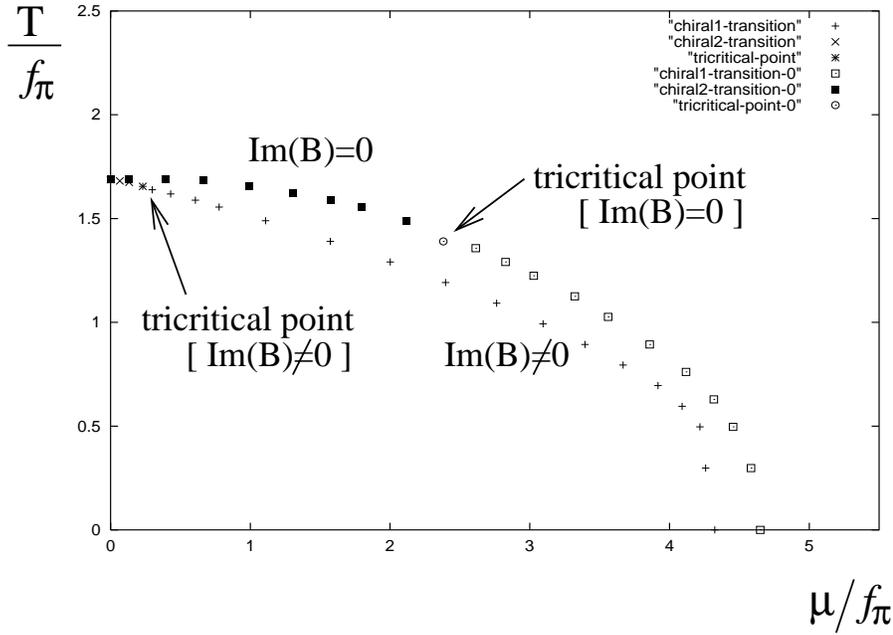}}
 \caption{Phase diagrams in two different cases 
 for $0\leq{T}/f_\pi\leq{2.5}$ and $0\leq\mu/f_\pi\leq{5.5}$. 
 The 2SC phase is not considered for simplicity.
 The lower-$T$ points show the chiral phase transiton points obtained from
 the SDE with ${\rm Im}[B]$, while the higher-$T$ points show those
 obtained from the SDE without ${\rm Im}[B]$. $f_\pi$ is 88 \mbox{MeV}.
}
 \label{pha-3case2}
\end{figure}

The value of $\mu$ at the tricritical point (end point)
obtained in Fig.~\ref{phase-structure} is much smaller than those in several 
other models, such as the 4-Fermi interaction model~\cite{Be,Asakawa:bq}
as well as in the other SDE analysis,~\cite{Bar,Kir1}
but is consistent with that obtained in Ref.~\citen{Ha},
which was also obtained through the analysis of the SDE.
We find that in the framework of the SDE,
the large difference in the position of the tricritical
point in the $T$-$\mu$ plane is
caused by the existence of the imaginary part of the Dirac mass
$B$ at $\mu\neq{0}$~\cite{takagi}.
See Fig.~\ref{pha-3case2}, in which the chiral phase transition points
are shown for two cases: one is the SDE analysis with ${\rm Im}[B]\neq{0}$
and the other is that with ${\rm Im}[B]={0}$.
When we use the SDE without the imaginary
part of the Dirac mass (i.e., ${\rm Im}[B]=0$),
the tricritical point is at $({T}, \mu)=(124, 210) \mbox{MeV}$.
The value of $\mu$ at this tricritical point [$\mu\sim{O(100) \mbox{MeV}}$]  
is close to the values 
obtained from analyses carried out using models 
with the local 4-Fermi interaction~\cite{Be,Asakawa:bq} 
and those carried out using the SDEs without Im[B]~\cite{Bar,Kir1}. 

In Ref.~\citen{takagi} we performed analysis 
with the imaginary part of the Dirac mass included, 
because in the SDE at non-zero chemical potential, 
the imaginary part ${\rm Im}[B]$, 
is inevitably 
generated in the hadron phase~\cite{takagi,Taka}.
However, some analyses of the SDE do not include   
the imaginary part~\cite{Bar,Kir1}, and analyses done using the local 4-Fermi 
interaction model do not generally include the imaginary part, 
because a leading order approximation is used.

\section{Summary and discussion}
\label{Summary and Discussion}

We studied the phase structure of hot and/or dense QCD(with massless 
2 flavors) by solving the Schwinger-Dyson equations 
for the Dirac and Majorana masses
of the quark propagator with the improved ladder approximation in the
Landau gauge.
There are two notable features: One is the position of the
tricritical point ( $\mu\sim{20}\mbox{MeV}$ ) and the other is the rather
large critical temperature ( $T\sim{100}\mbox{MeV}$ ) 
at the color phase transition. We find that the antiquark mass is 
of the same order as the quark mass in the low and medium density region,
and setting $\Delta^+=\Delta^-$ is actually a good approximation 
for investigating the phase diagram, the quark Majorana mass gap 
and the diquark condensate.  

We emphasize the importance of the imaginary part of
the Dirac mass at nonzero chemical potential.
We perform the analysis with the imaginary part of the Dirac mass 
${\rm Im}[B]$ included,
because the imaginary part is inevitably generated in the chiral symmetry 
breaking phase in the SDE at non-zero $\mu$.
We find that the most noteworthy feature 
of the analysis that includes the imaginary part of the Dirac mass 
is in the position of the tricritical point on the $T$-$\mu$ plane.
In the SDE analysis, including the imaginary part causes the tricritical point 
to move to a position of much smaller $\mu$: 
$({T}, \mu)=(124, 210) \mbox{MeV} \rightarrow ({T}, \mu)=(146, 20) \mbox{MeV}$.

\section*{Acknowledgements}

The author would like to thank D.~K.~Hong for 
informing his work~\cite{Hong:2000ng}.



\end{document}